\documentclass[a4paper]{article}

\usepackage{INTERSPEECH2019}
\usepackage{makecell}
\usepackage{array, xcolor}
\usepackage{subcaption}

\title{UWB-NTIS Speaker Diarization System for the DIHARD II 2019 Challenge}
\name{Zbyn\v{e}k Zaj\'ic$^1$, Marie Kune\v{s}ov\'a$^{1,2}$, Marek Hr\'uz$^{1}$, Jan Van\v{e}k$^1$,}
\address{
University of West Bohemia \\
Faculty of Applied Sciences \\
$^1$NTIS - New Technologies for the Information Society and $^2$Dept. of Cybernetics,\\
Univerzitn\'i\ 8, 306 14 Plze\v{n}, Czech Republic}
\email{\{zzajic, mkunes, mhruz, vanekyj\}@ntis.zcu.cz}
\begin{document}
\maketitle
\begin{abstract} 
In this paper, we present our system developed by the team from the New Technologies for the Information Society (NTIS) research center of the University of West Bohemia in Pilsen, for the Second DIHARD Speech Diarization Challenge. The base of our system follows the currently-standard approach of segmentation, i/x-vector extraction, clustering, and resegmentation. The hyperparameters for each of the subsystems were selected according to the domain classifier trained on the development set of DIHARD II. We compared our system with results from the Kaldi diarization (with i/x-vectors) and combined these systems. At the time of writing of this abstract, our best submission achieved a DER of 23.47\% and a JER of 48.99\% on the evaluation set (in Track 1 using reference SAD).
\end{abstract}


\section{Introduction}
In recent years, we have developed our Speaker Diarization (SD) system~\cite{ZZ_SD_Specom_2016, 2017_ICASSP_Hruz_Zajic, IS2017_SD_CNNvaha}. Last year, we also tailored an off-line system for the First DIHARD Speech Diarization Challenge~\cite{IS18_Dihard_Zajic}, where we participated in Track 1 as well as Track 2 of the challenge.  

The Second DIHARD Challenge~\cite{DIHARD2_plan} brought us an opportunity to extend our system and to try combining results from different sources. Besides i-vector~\cite{Kenny_2008_SpeakerVariability,Machlica_FA} extraction, we have also extracted x-vectors~\cite{Snyder2018}. The main novelty in our system is an early-fusion of i-vectors and x-vectors into xi-vectors and modified segmentation. We also use the Kaldi system\footnote{http://kaldi-asr.org/} with a recipe for diarization as an additional system for comparison and for a domain-specific combination of results. This year we have decided to participate only in the Track 1 part of the challenge, where a reference speech labeling is available and no speech activity detection (SAD) step is needed.

Our main advantage in the First DIHARD Challenge was the application of a Neural Network (NN) based domain classifier that allows the system to automatically identify the domain of each recording and to set the system's configuration accordingly. The same applies for the Second DIHARD Challenge.


\section{Speaker Diarization System}
\label{sec:system}

\begin{figure}[ht]
	\centering
	\includegraphics[scale=0.7]{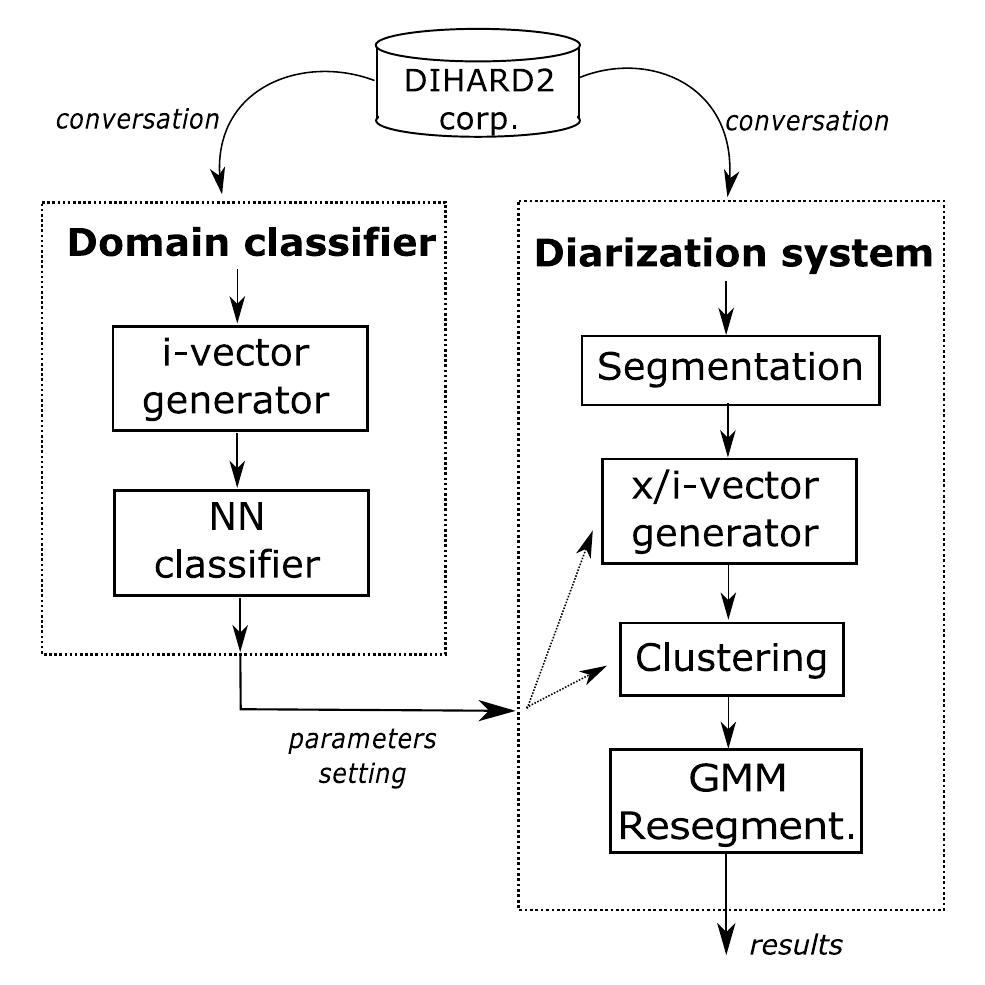}
	\caption{Diagram of the diarization process.} \label{fig:schema}
\end{figure}

Our system follows x-vector- and i-vector-based approaches~\cite{SD_iVec_PLDA_2014,senoussaoui2014study,Snyder2018}.
A diagram of our diarization system is shown in~Figure~\ref{fig:schema}. The general structure is the same as in our previous system for the First DIHARD Challenge~\cite{IS18_Dihard_Zajic}.



%

This section provides a description of the main steps of the diarization process. The domain classifier and related domain-dependent settings are described in section~\ref{sec:IdCorp}.

\subsection{Segmentation}
\label{subsec:segmentation}
First, the entire conversation is split into multiple individual speech regions by breaking it on any non-speech events; these non-speech regions are excluded from subsequent processing. As a second step, the speech regions are further segmented according to the probability of speaker change given by our Speaker Change Detector, which is based on a Convolutional Neural Network (SCD-CNN)~\cite{2017_ICASSP_Hruz_Zajic}. 

The SCD-CNN was trained as a regressor on spectrograms of the acoustic signal. The process is described in our previous paper for the First DIHARD Challenge~\cite{IS18_Dihard_Zajic}. The signal between two detected speaker changes is considered to be one segment.

To ensure that each segment contains sufficient information about the speaker, we set the minimum duration of each segment to 0.5\,s. Shorter segments are discarded from the clustering stage and the decision about the speaker is left for the resegmentation step (see section~\ref{subsec:resegm}).

\subsection{Feature Extraction} 
\label{subsec:features}

We use the same signal processing pipeline as for the First DIHARD challenge~\cite{IS18_Dihard_Zajic} -- Linear Frequency Cepstral Coefficients (LFCCs). As a newly added step, Cepstral Mean Normalization (CMN) is applied to compensate for channel variations.

\subsection{Segment Description}
\label{subsec:iVec}

Each segment is represented by a concatenation of the x-vector and i-vector for the same segment. We call this an xi-vector. The i- and x-vectors were obtained using a Kaldi recipe\footnote{https://github.com/kaldi-asr/kaldi/tree/master/egs/\\callhome\_diarization/v1 and /v2}~\cite{IS18_dihard_JHU}.
To exploit the ability of x/i-vectors to represent small amounts of data and to minimize the presence of more than one speaker in a segment, longer segments are split into intervals of max. 2\,s, with 1\,s overlaps.

For generating i-vectors, we trained a UBM with $2048$ components and a transformation matrix with a latent dimension of $128$. A Time Delay Neural Network is used as an x-vector extractor, and x-vectors are extracted from the affine component of the second-to-last layer with dimension $128$. 

For whitening the xi-vectors, we subtract the mean of development set's xi-vectors. During the diarization process, we use a conversation-dependent Principal Component Analysis (PCA)~\cite{Intra-ConversationVariability_SD_2011} computed on the data in the current conversation to reduce the dimension of the xi-vectors into $3$, $6$, $9$ or $12$  (depending on the identified corpus -- see Tab.~\ref{tab:parameters}). 

The following corpora were used as training data: LibriSpeech\footnote{http://www.openslr.org/12/}, VoxCeleb2~\cite{IS18_VoxCeleb2}, TedLium3~\cite{Specom18_TedLium3}, and the following ELRA corpora: Speecon database UK English (ELRA-S0215) and US English (ELRA-S0233). Additional data augmentation (additive noise, music, babble and reverberation) was used on the LibriSpeech and TedLium3 corpora.

\subsection{Clustering}
\label{subsec:cluster}


The general clustering approach remains mostly unchanged from what we used in the First DIHARD Challenge. 

As in the previous year's challenge, the number of speakers in each recording is unknown in advance: the DIHARD corpus consists of several distinct domains where the number of speakers ranges from 1 to 10. Thus, we have chosen to primarily use the agglomerative hierarchical clustering (AHC) algorithm.

The clustering process is based on the average cosine distance between xi-vectors. We use a fixed distance threshold as the stopping condition. Additionally, we leverage our knowledge from the development set by also specifying a minimum and maximum number of clusters for each conversation, based on the number of speakers typically observed for the domain. So, we force the final number of clusters to be within this set range. The range and the clustering stopping condition -- the distance threshold -- were both established on a per-corpus basis using the development data (see Section \ref{sec:exper}).

In the Second DIHARD Challenge, there are three corpora in the development set with an overwhelming majority of two-speakers conversations. For these domains, we use a Probabilistic Linear Discriminant Analysis (PLDA) model~\cite{Ioffe2006_PLDA} to evaluate the distance between the xi-vectors. Then, we apply k-medoids clustering into a constant number of clusters across the whole domain. For training the PLDA model, we used the same datasets as listed in Subsection~\ref{subsec:iVec}. The between-class dimension is equal to the feature dimension.

\subsection{Resegmentation}
\label{subsec:resegm}
Finally, we refine the diarization results via resegmentation. The previous results are based on relatively long signal windows and the boundaries between speaker segments are not precise. Therefore, we compute a GMM for each speaker cluster from all feature vectors assigned to the speaker. Then, likelihoods for all speaker GMMs are evaluated and filtered by a Gaussian window (length 75\,ms with shift 50\,ms) to smooth peaks. The number of GMM components ranges between 1 and 64 depending on the speaker data size. Then, the entire conversation is reclassified according to the GMM likelihoods on a frame-by-frame basis.


\section{Domain Classification}
\label{sec:IdCorp}
The DIHARD II corpus~\cite{DIHARD2_corpus} consists of data taken from several different domains, with very diverse characteristics -- including the number of speakers, the level of noise, and audio quality in general. To improve the results of the diarization system, we decided to use the supervised information about each domain given by the organizers (the possible number of speakers in the domain) and to tune specific settings on development data (mainly the threshold for AHC).

We have proposed the domain classifier as a hierarchical two-stage classifier. The first level is a special classifier to distinguish recordings with one speaker from multi-speaker data. The second stage classifier is applied when the first level class is considered to be the multiple speaker case. It evaluates the posterior probability that the input conversation belongs to one of the 11 corpora in the DIHARD II development set.

The same NN architecture is used for both classifiers. They differ only in the last layer, where the first level NN uses one neuron for the binary classification, and the second stage NN uses a softmax layer with 11 classes.

The NNs receive a single i-vector calculated over the entire conversation as the input. A special i-vector extractor was trained for the domain classifiers. The LFCC features are the same but also include frames marked as non-speech by SAD. The UBM has 512 diagonal components, and the final i-vector dimension is 100.

The NNs have one hidden layer with 2048 neurons and tanh activation function. Dropout with coefficient 0.9 was used during training. The network was implemented in TensorFlow, 
 where the ``adam'' optimizer was used with 10 epochs and batch size of 32. The remaining hyper-parameters were left at default values.

Both NNs were trained on the development data + 10 randomly chosen recordings from the LibriSpeech corpus (cut to 10min length).

Because the evaluation dataset contains two unknown corpora with very different characteristics, we apply a threshold on the classifier result. The positive detection threshold was set to 0.6 for both stage classifiers. If there is no positive detection in the second-stage NN, the conversation is treated as ``unknown domain''. The accuracy of the first-stage NN was 100\% on a held-out part of the development data. The accuracy of the second-stage NN was 82\%.

\subsection{Domain-specific settings}


Because of our domain classifier, we were able to use different system configuration for each of the 11 development set corpora and for unknown data. Here we describe the general approaches we selected for each domain. Specific experimentally-chosen parameters are listed in Table~\ref{tab:parameters}.

 The descriptions of the individual DIHARD II corpora can be found in the challenge evaluation plan~\cite{DIHARD2_plan} as well as in the main challenge paper~\cite{DIHARD2_Ryant2019}. As such, we do not replicate them here. 
 
    \emph{LibriVox:} All recordings contain only 1 speaker. Thus, we did not need to perform diarization but simply used the information given by reference SAD. We rely on the first stage of the domain specific classifier.
    
    \emph{SEEDLingS, SCOTUS, RT-04S, SLX, VAST and YouthPoint:} For these corpora, we used the AHC approach with cosine distance, as described in section \ref{subsec:cluster}.
    
    \emph{ADOS, DCIEM and MIXER6:} These corpora have almost exclusively exactly 2 speakers in each conversation. For this reason, we could simply use k-medoids clustering into 2 clusters with PLDA scores. 
    
    \emph{CIR:} For this corpus, our system gives the best results on development data for k-medoids clustering into 4 clusters with PLDA scores.

    \emph{Unknown:} For unrecognized evaluation data, we've chosen to use AHC with 2-6 target clusters.
    
    \begin{table}[ht] 
    	\centering
    	\caption{Experimentally chosen parameters (Thr. = threshold, k-m = k-medoids) for each corpus.}	\begin{tabular}{lccccc}
    		\toprule
    		\textbf{Corpus}  & \textbf{\thead{Clustering}} & \textbf{\thead{No. \\ spk}}  &  \textbf{\thead{Thr \\ AHC}} &  \textbf{\thead{PCA \\ dim}}\\
    		\midrule
    		{LibriVox}  & - & 1 & - & -\\
    		{SEEDL.}	& AHC & 2-3  & 0.62 & 6\\
    		{CIR}	    & k-m & 4  & - & -\\
    		{ADOS}     	& k-m & 2 & - & -\\
    		{SCOTUS}  	& AHC & 5-10 & 0.46 & 12\\
    		{DCIEM}     & k-m & 2 & - & -\\
    		{RT-04S}    & AHC & 3-10 & 0.46 & 6\\
    		{SLX}     	& AHC & 2-6 & 0.762 & 6\\
    		{MIXER6}    & k-m & 2 & - & -\\
    		{VAST}     	& AHC & 1-9 & 0.58 & 3\\
    		{YouthP.}  	& AHC & 3-5 & 0.54 & 9\\
    		{other}    	& AHC & 2-6  & 0.1 & -\\
    		\bottomrule
    	\end{tabular}
    	\label{tab:parameters}
    \end{table}
    \noindent

\section{Kaldi Diarization System}
\label{sec:kaldi_system}
As an additional system, we have decided to use a Kaldi recipe for diarization~\cite{IS18_dihard_JHU}. The input features are the same LFCCs as in our system (details in Section~\ref{subsec:features}).

The segmentation provides chunks of speech between important non-speech events (Kaldi SAD segmentation) and subsequently divides these segments into sub-segments with constant length 1.5\,s and overlap 0.75\,s (the minimum length of a segment is 0.5\,s).

X-vectors or i-vectors are computed on segmented data and handled by a PLDA model to compute the similarity between these segments. X/i-vectors are whitened before the PLDA estimation by subtracting the mean and transforming by an LDA matrix.

The vectors of segments are then clustered according to the AHC, with the stopping threshold set on development data. This threshold was found for the entire development set -- this system does not treat different domains of the DIHARD II corpus differently.

Additionally, we create xi-vectors by concatenating the x-vector and i-vector for the same segment. The whitening transformation is also obtained by concatenating the means and PCA transformation matrices belonging to the x/i-vectors (this independent treatment works better than computing the whitening transformation on the whole xi-vectors).

\section{Late system combination}
\label{sec:mix_system}
The Kaldi diarization system does not use the information from the domain classifier, and its setting is very general. Therefore, we have used the Kaldi system instead of our speaker diarization system (SD) in the cases where the domain classifier marked the conversation as ``unknown domain''.

For the two most problematic corpora (Seedlings, VAST), we have also used the Kaldi system. On average, our system slightly outperforms the Kaldi system on the development data for these corpora. However, the DERs of individual conversations have a higher variance than the ones from Kaldi. We refer to this per-domain system selection as late system combination.

\section{Experiments}
\label{sec:exper}

This section describes our experiments on the development set of the Second DIHARD Challenge, as well as our final results on the evaluation set. The experiments mainly served for finding the optimal system configuration for each of the individual corpora. For details of the DIHARD II corpus~\cite{DIHARD2_corpus, SEEDLINGS_corpus}, see the evaluation plan~\cite{DIHARD2_plan}.

\subsection{Evaluation}
\label{subsec:evaluation}

The system performance was evaluated in terms of Diarization Error Rate (DER), as defined by NIST~\cite{DER_Nist2006}. On the development set, we calculated this on a per-recording basis using NIST's md-eval.pl script\footnote{https://github.com/usnistgov/SCTK/blob/master/src/md-eval}. 

DER and Jaccard Error Rate (JER) on the evaluation set were given by the official scoring system~\cite{DIHARD2_Ryant2019}. 

Unlike usual practice, DIHARD Challenge submissions were scored with no forgiveness collar around speaker boundaries, and overlapping speech was included in the evaluation.

\subsection{Results}
\label{sec:res}
Table~\ref{tab:resultsDEV_wholeSys} presents a comparison between i/x/xi-vectors contribution with an earlier version of our system. Based on these preliminary results, the rest of our work was with xi-vectors only.

Table~\ref{tab:resultsDEV_IcCorp} shows results on the development set for each of the eleven corpora. Table~\ref{tab:resultsTEST} then presents the final results on the evaluation data for Track 1 -- diarization using reference SAD.

\begin{table}[ht]
  \centering
  \caption{Average DER~[\%] on DIHARD II development set for an earlier version of our system and for Kaldi with different segment descriptors (x/i/xi-vector).}
    \begin{tabular}{lcc}
    \toprule
    \textbf{system} & \textbf{DER}\\
    \midrule
    {SD i-vec}  			& 24.31 \\
    {SD x-vec}     		    & 23.81\\
    {SD xi-vec}     		& \textbf{22.51}\\
    {Kaldi i-vec}  			& 25.83\\
    {Kaldi x-vec}     		& 25.32\\
    {Kaldi xi-vec}     		& 24.13\\
    \bottomrule
    \end{tabular}
  \label{tab:resultsDEV_wholeSys}
\end{table}
\noindent

\begin{table}[ht]
  \centering
  \caption{Average DER~[\%] on individual corpora of the DIHARD II development set, for our system (SD), Kaldi system, and the combination system, all using xi-vectors.}
    \begin{tabular}{lccc}
    \toprule
    \textbf{Corpus} & \textbf{SD} & \textbf{Kaldi} & \textbf{Comb.}\\
    \midrule
    {LibriVox}     		& 0.00 & 14.52 & 0.0 \\
    {SEEDLingS}			& 31.32 & 33.90 & 33.90 \\
    {CIR}			    & 45.83 & 52.25 & 45.83  \\
    {ADOS}     			& 14.06 & 16.01 & 14.06 \\
    {SCOTUS}  			& 6.92 & 18.03 & 6.92 \\
    {DCIEM}     		& 8.88 & 9.65 & 8.88 \\
    {RT-04S}     		& 33.14 & 36.30 & 33.14 \\
    {SLX}     			& 17.56 & 16.90 & 17.56 \\
    {MIXER6}            & 9.42 & 9.72 & 9.42 \\
    {VAST}     			& 38.00 & 39.65 & 39.65 \\
    {YouthPoint}     	& 4.55 & 6.33 & 4.55 \\
    \midrule   
    {All}				& \textbf{20.78} & 24.13  & 21.29\\
    \bottomrule
    \end{tabular}
  \label{tab:resultsDEV_IcCorp}
\end{table}
\noindent

\begin{table}[ht]
  \centering
  \caption{Official results (DER~[\%] and JER~[\%]) on the DIHARD II evaluation data for our system (SD), Kaldi, and Comb.}
    \begin{tabular}{lccc}
    \toprule
    \textbf{} & \textbf{SD} & \textbf{Kaldi} & \textbf{Comb.}\\
    \midrule
         \thead{DER[\%] \\ JER[\%]} 	& \thead{24.59\\49.63} & \thead{25.17\\54.94} & \thead{\textbf{23.47}\\\textbf{48.99}} \\
   \bottomrule    
     \end{tabular}
  \label{tab:resultsTEST}
\end{table}
\noindent

\section{Discussion}
\label{sec:discuss}

For this challenge, we decided to extend our previous system for diarization with various enhancements; this chapter discusses their benefits. Table~\ref{tab:resultsDEV_IcCorp_Discuss} presents the results for our SD system with comparison with several alterations: a system without SCD-CNN (only Kaldi SAD segmentation), with de-noised test data (using speech enhancement\footnote{https://github.com/staplesinLA/denoising\_DIHARD18}~\cite{IS18_dihard_LeiSun}, no de-noising for training data), and a system with reference information about overlapping speech. The latter information is used in the resegmentation step -- for such parts of the data, the second most likely speaker is also detected.

\begin{table}[ht]
  \centering
  \caption{Average DER~[\%] on individual corpora of the DIHARD II development set, for our system (SD) with different setting -- with Kaldi SAD segmentation instead of SCD-CNN, test data de-noise and with reference overlap labels.}
    \begin{tabular}{lcccc}
    \toprule
    \textbf{Corpus} & \textbf{SD} & \textbf{\thead{without \\ SCD}}  & \textbf{\thead{with \\ de-noise}}  & \textbf{\thead{with ref.\\ overlap}}\\
    \midrule
    {LibriVox}     		& 0.00 & 0.0 & 0.0 & 0.0\\
    {SEEDLingS}			& 31.32 & 31.22 & 32.30 & 24.56\\
    {CIR}			    & 45.83 & 47.88 & 46.70 & 37.71\\
    {ADOS}     			& 14.06 & 13.26 & 14.25 & 10.73\\
    {SCOTUS}  			& 6.92 & 10.67 & 8.01 & 5.99\\
    {DCIEM}     		& 8.88 & 8.66 & 8.74 & 6.24\\
    {RT-04S}     		& 33.14 & 36.38 & 34.53 & 25.69\\
    {SLX}     			& 17.56 & 19.14 & 17.36 & 13.64\\
    {MIXER6}            & 9.42 & 9.29 & 9.93 & 5.02\\
    {VAST}     			& 38.00 & 38.91 & 38.61 & 30.09\\
    {YouthPoint}     	& 4.55 & 5.26 & 5.49 & 3.89\\
    \midrule   
    {All}				& 20.78 & 21.52  & 21.31 & 16.16\\
    \bottomrule
    \end{tabular}
  \label{tab:resultsDEV_IcCorp_Discuss}
\end{table}
\noindent

Table~\ref{tab:coveragePurity} shows the segmentation coverage and purity~\cite{ICASSP2017_TripletLossSR} achieved on the development set for SCD-CNN and Kaldi SAD segmentation. Our SCD-CNN method outperforms the Kaldi SAD segmentation in both coverage and purity, except two cases. This result is expected, as SAD segmentation does not take into account the speaker changes in one speech activity segment. 
The purity measure is more important in the diarization task, as it tells us how pure the segments are -- that they contain only one speaker. 

\begin{table}[ht]
  \centering
  \caption{Overall Coverage and Purity for segments provided by Kaldi (SAD) and by SCD-CNN}
  \begin{subtable}{0.35\linewidth}
  \caption{SAD segments}
    \begin{tabular}{lcc}
    \toprule
    \textbf{Corpus} & \textbf{Cov} & \textbf{Pur}\\
    \midrule
    VAST      & \textbf{0.945} & 0.454 \\
    YP        & 0.950 & 0.445 \\
    LIBRIVOX  & 0.948 & 0.466 \\
    SEEDLINGS & 0.917 & 0.600 \\
    CIR       & 0.963 & 0.345 \\
    ADOS      & 0.896 & 0.669 \\
    SCOTUS    & 0.980 & 0.375 \\
    DCIEM     & 0.901 & 0.563 \\
    RT04      & \textbf{0.969} & 0.421 \\
    SLX       & 0.916 & 0.507 \\
    MIXER6    & 0.933 & 0.440 \\
    \bottomrule
    \end{tabular}
  \end{subtable}
  \hspace{0.28\linewidth}
  \begin{subtable}{0.3\linewidth}
  \caption{SCD-CNN}
    \begin{tabular}{cc}
    \toprule
    \textbf{Cov} & \textbf{Pur}\\
    \midrule
    0.926 & \textbf{0.820} \\
    \textbf{0.991} & \textbf{0.971} \\
    \textbf{0.987} & \textbf{0.991} \\
    \textbf{0.937} & \textbf{0.868} \\
    \textbf{0.979} & \textbf{0.588} \\
    \textbf{0.992} & \textbf{0.875} \\
    \textbf{0.990} & \textbf{0.936} \\
    \textbf{0.987} & \textbf{0.862} \\
    0.889 & \textbf{0.670} \\
    \textbf{0.988} & \textbf{0.871} \\
    \textbf{0.958} & \textbf{0.868} \\
    \bottomrule
    \end{tabular}
  \end{subtable}
  \label{tab:coveragePurity}
\end{table}

Based on these results, we chose to apply the SCD-CNN in our final system, and we omitted the de-noising. Unfortunately, despite the clear benefits of detecting overlapping speech, we were not able to train a real overlap detector with reasonable accuracy on the DIHARD II data.

\section{Conclusion}
\label{sec:concl}
In this paper, we presented a new version of our diarization system and its results for the Second DIHARD Diarization Challenge. Compared to our previous system, we applied xi-vectors and modified the SCD-based segmentation step to take advantage of x/i-vectors' ability to represent short segments. Using a domain classifier as in the previous challenge, we were able to use a different system configuration for each subset of data. For comparison, we applied a Kaldi recipe for diarization and combined the results into a single system. Additionally, we have investigated the potential gains of detecting overlapping speech, de-noising, and various segmentation methods. Our best Track 1 submission achieved a DER of 23.47\% and JER of 48.99\%.

\section{Acknowledgements}
This research was supported by the Ministry of Culture of the Czech Republic, project No.DG16P02B048. Access to computing and storage facilities provided by the project CESNET LM2015042 is greatly appreciated.



\bibliographystyle{IEEEtran}
\bibliography{library_no_url}


\end{document}